\documentclass[aps,twocolumn,preprintnumbers,amsmath,amssymb,floatfix,nofootinbib]{revtex4-1}
\usepackage{graphicx}

\usepackage{color}
\usepackage{colordvi}
\usepackage[dvipsnames]{xcolor}
\usepackage{slashed}
\usepackage{ulem}

\usepackage{hyperref}

\usepackage{xcolor}

\newcommand{\beqn}{\begin{eqnarray}}
\newcommand{\eeqn}{\end{eqnarray}}
\newcommand{\be}{\begin{equation}}
\newcommand{\ee}{\end{equation}}

\newcommand{\ba}{\begin{array}{c}}
\newcommand{\bat}{\begin{array}{cc}}
\newcommand{\ea}{\end{array}}

\newcommand{\bi}{\begin{itemize}}
\newcommand{\ei}{\end{itemize}}

\DeclareMathOperator{\dif}{d\!}
\DeclareMathOperator{\br}{Br}

\usepackage{cleveref}
\crefname{equation}{Eq.}{Eqs.}
\crefname{figure}{Fig.}{Figs.}
\crefname{table}{Table}{Tables}
\crefname{Section}{Section}{Sections}
\allowdisplaybreaks

\begin{document}

\preprint{LMU-ASC 39/16}

\preprint{TUM-HEP-1069/16}


\title{Dirac dark matter and $b \to s  \ell^+ \ell^-$ with $\mathrm{U(1)}$ gauge symmetry}

\author{Alejandro Celis${}^{1}$}
\author{Wan-Zhe Feng${}^{2}$}
\author{Martin Vollmann${}^{3}$}

\affiliation{${}^1$ Ludwig-Maximilians-Universit\"at M\"unchen,
   Fakult\"at f\"ur Physik,\\
   Arnold Sommerfeld Center for Theoretical Physics,
   80333 M\"unchen, Germany}
\affiliation{${}^2$  Department of Physics, Northeastern University, Boston, MA 02115-5000, USA}
\affiliation{${}^3$ Physik Department T31. James-Franck-Stra\ss{}e 1,\\ Technische Universit\"at M\"unchen, 85748 Garching, Germany}

\begin{abstract}

We revisit the possibility of a Dirac fermion dark matter candidate in the light of current $b \to s  \ell^+ \ell^-$ anomalies by investigating a minimal extension of the Standard Model with a horizontal $\mathrm{U(1)}^{\prime}$ local symmetry.  Dark matter stability is protected by a remnant $\mathcal{Z}_2$ symmetry arising after spontaneous symmetry breaking of $\mathrm{U(1)}^{\prime}$.    The associated  $Z^{\prime}$ gauge boson can accommodate current hints of new physics in $b \to s  \ell^+ \ell^-$ decays, and acts as a vector portal between dark matter and the visible sector.   We find that the model is severely constrained by a combination of precision measurements at flavour factories,  LHC searches for dilepton resonances, as well as direct and indirect dark matter searches.    Despite this, viable regions of the parameter space accommodating the observed dark matter relic abundance and the $b \to s  \ell^+ \ell^-$ anomalies still persist for dark matter and $Z^{\prime}$ masses in the TeV range.

\end{abstract}

\maketitle

\vspace{-5cm}

\section{Introduction}

A series of deviations from the Standard Model (SM) of particle physics have been observed recently in semi-leptonic $b \to s \ell^+ \ell^-$ decays.  Notably, departures from lepton universality at the $25\%$ level ($2.6\sigma$ deviation from the SM) have been reported by the LHCb collaboration in $R_K = \Gamma(B^+ \to K^+ \mu^+ \mu^-)/\Gamma(B^+ \to K^+ e^+ e^-)$~\cite{Aaij:2014ora}.  This observable is predicted within the SM with a high precision given that many sources of uncertainties cancel when taking the ratio~\cite{Hiller:2003js,Guevara:2015pza,Bordone:2016gaq}.

Additionally, the LHCb collaboration has also reported deviations from the SM in exclusive $b \to s \mu^+ \mu^-$ decay observables such as branching fractions and angular distributions.     The latest global fits of $b \to s \mu^+ \mu^-$ data favor new physics contributions in the effective operator $(\bar s \gamma_{\mu}   P_L b) ( \bar \mu \gamma^{\mu} \mu)$ of about $\sim 20\%$ of the SM contribution~\cite{Alonso:2014csa,Ghosh:2014awa,Jager:2014rwa,Altmannshofer:2015sma,Descotes-Genon:2015uva,Ciuchini:2015qxb,Hurth:2016fbr}.  Interestingly, new physics contributions of this type also accommodate the hint for lepton universality violation in $R_K$.

These observations have motivated the consideration of $\mathrm{U(1)}^{\prime}$ gauge extensions of the SM~\cite{Altmannshofer:2014cfa,Crivellin:2015mga,Crivellin:2015lwa,Celis:2015ara,Falkowski:2015zwa,Celis:2015eqs,Sierra:2015fma,Allanach:2015gkd,Belanger:2015nma}.    A $Z^{\prime}$ boson arising from a spontaneously broken $\mathrm{U(1)}^{\prime}$ symmetry can mediate $b \to s \ell^+ \ell^-$ transitions at tree-level producing significant deviations from the SM.       These models resort to family non-universal $\mathrm{U(1)}^{\prime}$ charge assignments and/or fermion mixing effects with additional heavy fermions in order to generate the required flavour changing $Z^{\prime}$ couplings to quarks and lepton non-universality.\footnote{The possibility that such $Z^{\prime}$ boson arises from the spontaneous breaking of a non-abelian gauge group has been discussed in~\cite{Chiang:2016qov,Boucenna:2016wpr,Boucenna:2016qad}.   }

While considering extensions of the SM with a local $\mathrm{U(1)}^{\prime}$ symmetry motivated by $b \to s \ell^+ \ell^-$ decay anomalies, it is tempting to ponder on the possibility that such symmetry is also behind the observed dark matter (DM) in our universe.    Previous work in this direction has been done initially in~\cite{Sierra:2015fma} and later followed in~\cite{Belanger:2015nma}.  These works explored a scalar DM candidate which is stabilized by a remnant $\mathcal{Z}_2$ global symmetry appearing after spontaneous symmetry breaking.   We discuss in this work the possibility of a Dirac fermion DM candidate which is subject to a similar protection mechanism.  We do so by examining in detail a minimal extension of the SM.

A  Dirac fermion charged only under a new $\mathrm{U(1)}^{\prime}$ gauge symmetry is a popular DM candidate considered in the literature~\cite{Alves:2013tqa,Duerr:2013lka,Jackson:2013pjq,Jackson:2013rqp,Arcadi:2013qia,Duerr:2014wra,Bell:2014tta,Feng:2014cla,Hooper:2014fda,Alves:2015pea,Duerr:2015wfa,Alves:2015mua}.   We explore this idea in the light of the current $b \to s \ell^+ \ell^-$ data.  More generally, we discuss possible connections between DM searches and flavour physics experiments arising in the context of flavoured $\mathrm{U(1)}^{\prime}$ extensions of the SM.

The choice of the $\mathrm{U(1)}^{\prime}$ symmetry is mostly restricted by the requirement of generating sizable effects in $b \to s \ell^+ \ell^-$ transitions.  We choose for concreteness a horizontal $\mathrm{U(1)}^{\prime}$ symmetry under which the quark fields transform non-trivially, corresponding to the combination of family baryon numbers considered in~\cite{Crivellin:2015lwa}.   Flavour violating $Z^{\prime}$ couplings to down-type quarks are generated at tree-level and are controlled to a good approximation by elements of the quark-mixing matrix.       The $Z^{\prime}$ couplings to leptons will be generated via fermion mixing effects with additional vector-like fermions.

In our scenario, DM annihilation into SM particles proceeds through the $Z^{\prime}$ gauge boson.  Accommodating the observed $b \to s \ell^+ \ell^-$ anomalies while satisfying current flavour and collider constraints restricts the $Z^{\prime}$ mass to be around the TeV scale.     The observed value of the DM relic density can be obtained via the Wigner enhancement effect of DM annihilation,
in which case the mass of DM candidate is around half of the $Z^{\prime}$ mass.
Thus, our model predicts a TeV scale DM when current $b \to s \ell^+ \ell^-$ data is taken at face value.

We review the extension of the SM with a gauged family baryon number symmetry in Sec.~\ref{sec:gbn}, which forms the basic framework used in our discussion.    The phenomenology of $B \to K^{(*)} \mu^+ \mu^-$ decays taking into account relevant flavour and collider bounds is discussed in Sec.~\ref{btok}.   We make the connection with dark matter in Sec.~\ref{sec:dmd} and present our conclusions in Sec.~\ref{sec:conc}.

\section{Gauged family baryon number} \label{sec:gbn}

We consider extending the SM by an anomaly free combination of family baryon numbers $\mathrm{U(1)}^{\prime} =  B_1 + B_2 -2 B_3$, where $B_{i}$ denotes the baryon number of the $i$th family~\cite{Crivellin:2015lwa}.          The charge assignments of the quark fields under the $\mathrm{U(1)}^{\prime}$ symmetry are
\begin{align}
X_{Q_L} &= \left(  - \frac{1}{3} , - \frac{1}{3}, \frac{2}{3} \right) \,, \quad
X_{u_R} = X_{d_R}  = \left( - \frac{1}{3} ,  - \frac{1}{3}, \frac{2}{3} \right)  \,.
\end{align}
In order to obtain quark masses and mixing we include a second Higgs doublet to the SM field content.     We consider then two complex Higgs doublets with $\mathrm{U(1)}^{\prime}$ charge
\begin{align}
X_{H_1} =  -1  \,, \qquad  X_{H_2} = 0\,.
\end{align}
Charged leptons only have Yukawa couplings with $H_2$ due to the charge assignments.     The electrically neutral component of the Higgs doublets acquire non-vanishing vacuum expectation values (vevs) breaking the electroweak symmetry, $\langle H_i^0 \rangle = v_i/\sqrt{2}$ with  $v^2 = v_1^2 + v_2^2 =   (\sqrt{2} G_F )^{-1}$.      Additionally, a complex scalar field $\phi$ that is singlet under the SM gauge group and has $\mathrm{U(1)}^{\prime}$ charge is assumed to break spontaneously the $\mathrm{U(1)}^{\prime}$ symmetry around the TeV scale.

The $Z^{\prime}$ boson receives a TeV scale mass from the vev of $\phi$, $M_{Z^{\prime}} \sim   \sqrt{2} g^{\prime}  X \langle  \phi  \rangle $.  Here we have denoted the $\mathrm{U(1)}^{\prime}$ charge of $\phi$ as $X$.   In order to avoid an accidental $\mathrm{U(1)}$ global symmetry in the scalar potential and its corresponding Goldstone boson either $H_1^{\dag}  H_2  \phi^{(*)}$ or $H_1^{\dag}  H_2 ( \phi^{(*)})^2$ should be gauge invariant.  The $\mathrm{U(1)}^{\prime}$ charge of $\phi$ is then constrained to be $X= \pm1$ or $X =  \pm 1/2$. Details of the scalar potential are given in Appendix~\ref{aop:sec}.

Neglecting kinetic mixing among the two $\mathrm{U(1)}$ gauge factors,  the mixing between the two massive neutral vector bosons is described by~\cite{Babu:1997st}
\be  \label{emix}
\begin{pmatrix}
Z_\mu\\
Z^{\prime}_\mu
\end{pmatrix} = \begin{pmatrix}
c_\xi&s_\xi\\
-s_\xi&c_\xi
\end{pmatrix}
\begin{pmatrix}
\hat{Z}_\mu\\
\hat{Z}_\mu^\prime
\end{pmatrix}\,,\quad   \xi  \approx -\frac{e}{2 c_W s_W g^{\prime}}  \frac{  \langle H_1^0 \rangle^2     }{   X^2 \langle \phi \rangle^2  } \,.
\end{equation}
Here we have denoted $c_W = \cos \theta_W$ and $s_W = \sin \theta_W$ with $\theta_W$ being the weak mixing angle.  We are interested in the following in large $\tan \beta \equiv v_2/v_1$ values, $\tan \beta  \sim \mathcal{O}(10-10^2)$.      In this limit one avoids large new physics effects in $Z$-pole observables measured at LEP as the $Z- Z^{\prime}$ mass mixing is suppressed.

The Yukawa interactions in the model are given by
\begin{align}
- \mathcal{L}_Y = \bar Q \,\Gamma_i \, H_i P_R \, d + \bar Q \, \Delta_i  \, \tilde H_i P_R \,u + \bar L \, \Pi  \,H_2 P_R  \,\ell + \mathrm{h.c.}
\end{align}
with $P_{L,R}$ denoting the usual chirality projectors.  The lepton Yukawa matrix $\Pi$ is a generic complex matrix in family space while the quark Yukawa matrices have specific textures determined by the gauge symmetry:
%
\beqn \label{eq:BN}
\begin{aligned}
\Gamma_1&=
\begin{pmatrix}
0&0&\times\\
0&0&\times\\
0&0&0
\end{pmatrix}\,,\quad
\Gamma_2=
\begin{pmatrix}
\times & \times & 0\\
\times & \times & 0\\
0& 0& \times
\end{pmatrix}\,, \\[0.2cm]
\Delta_1&=
\begin{pmatrix}
0 &0 &0\\
0 &0&0\\
\times & \times &0
\end{pmatrix} \,,~\,
\Delta_2=
\begin{pmatrix}
\times & \times & 0\\
\times & \times & 0\\
0& 0& \times
\end{pmatrix} \,.
\end{aligned}
\eeqn
%
The fermion mass matrices read
\begin{align}
M_d &= \frac{v}{ \sqrt{2} }  \left(  c_{\beta} \, \Gamma_1    +  s_{\beta} \,  \Gamma_2 \right) \,, \nonumber \\
M_u &= \frac{v}{ \sqrt{2} }  \left(  c_{\beta} \, \Delta_1    +  s_{\beta} \,  \Delta_2 \right) \,, \nonumber \\
M_{\ell} &=   \frac{v}{  \sqrt{2}}   s_{\beta}   \Pi \,.
\end{align}
The fermion mass matrices are diagonalized via bi-unitary transformations
\begin{align}
M_d^{\rm{diag}}  = V_{dL}^{\dag} \, M_d  \, V_{dR}  \,, \quad M_u^{\rm{diag}}  = V_{uL}^{\dag} \, M_u  \, V_{uR}  \,,
\end{align}
and similarly for the lepton sector $M_{\ell}^{\rm{diag}}  = V_{\ell L}^{\dag} \, M_\ell  \, V_{\ell R}$.     The quark mixing matrix (CKM) is given by the product $V = V_{uL}^{\dag}  V_{dL}$.

Following~\cite{Crivellin:2015lwa} we consider the Yukawa couplings with the Higgs doublet  $H_1$ as small perturbations to the Yukawa structure.       Note that for large $\tan \beta$ values this is a natural limit to take.        A perturbative diagonalization of the quark mass matrices can be performed by expanding over small quark-mass ratios and the small perturbations $\Gamma_1, \Delta_1$.  Note that
\begin{align}
(M_q^{\rm{diag}})^2  =   V_{qL}^{\dag}  M_q^{\rm{diag}} M_q^{\rm{diag}  \dag}   V_{qL} =  V_{qR}^{\dag}  M_q^{\rm{diag}  \dag}  M_q^{\rm{diag}}   V_{qR}  \,,
\end{align}
for $q=u,d$.    In the up-quark sector we have
\begin{align}
 M_u^{\rm{diag}} M_u^{\rm{diag}  \dag}   &=  \begin{pmatrix}
0 & 0 & 0\\
0 & 0 & 0\\
0& 0& m_t^2
\end{pmatrix}   + \mathcal{O}(  m_{u,c}^2,  m_{u,c}  \tilde \Delta_1, \tilde  \Delta_1^2  ) \,, \nonumber \\
M_u^{\rm{diag}  \dag}   M_u^{\rm{diag}}  &=  m_t  \begin{pmatrix}
0 & 0 &  \tilde \Delta_{1,31}   \\
0 & 0 & \tilde  \Delta_{1,32}  \\
\tilde \Delta_{1,31}   & \tilde \Delta_{1,32}  & m_t
\end{pmatrix}  + \mathcal{O}(  m_{u,c}^2,  \tilde \Delta_1^2  ) \,.
\end{align}
Here we have denoted $\tilde \Delta_{1}  = v c_{\beta} \Delta_1/\sqrt{2}$.  This implies
\begin{align}
V_{uL} = \mathbf{1} +  \mathcal{O}(m_{u,c}/m_t,   \sqrt{m_{u,c} \tilde \Delta_1}/m_t,  \tilde \Delta_1/m_t  ) \,,
\end{align}
so that the CKM matrix is given by $V\simeq V_{dL}$.  In the down-quark sector we have
\begin{align}
 M_d^{\rm{diag}} M_d^{\rm{diag}  \dag}   &=    m_b  \begin{pmatrix}
0 & 0 & \tilde  \Gamma_{1,13}   \\
0 & 0 &\tilde  \Gamma_{1,23}  \\
\tilde \Gamma_{1,13}   & \tilde \Gamma_{1,23}  & m_b
\end{pmatrix}  + \mathcal{O}(  m_{d,s}^2, \tilde \Gamma_1^2  ) \,, \nonumber \\
 M_d^{\rm{diag}  \dag}   M_d^{\rm{diag}}  &= \begin{pmatrix}
0 & 0 & 0\\
0 & 0 & 0\\
0& 0& m_b^2
\end{pmatrix}   + \mathcal{O}(  m_{d,s}^2,  m_{d,s} \tilde \Gamma_1,  \tilde  \Gamma_1^2  ) \,,
\end{align}
with $\tilde \Gamma_{1}  = v c_{\beta} \Gamma_1/\sqrt{2}$.  This implies
\begin{align}
V_{dR} = \mathbf{1} +    \mathcal{O}(m_{d,s}/m_b,   \sqrt{m_{d,s} \tilde \Gamma_1}/m_b, \tilde \Gamma_1/m_b  ) \,.
\end{align}
The $Z^{\prime}$ therefore has flavour changing couplings to left-handed down-type quarks which are controlled by CKM matrix elements to a good approximation.     There are also right-handed flavour violating $Z^{\prime}$ coupling to up-type quarks, controlled by the matrix elements of $\Delta_1$.    We set $\Delta_1 = 0$ for simplicity given that we are not interested in flavour effects in the up-quark sector.  In this case both $V_{uL}$ and $V_{uR}$ commute with the charge matrices $\mathrm{diag}(X_{Q_L}), \mathrm{diag}(X_{u_R})$ and the $Z^{\prime}$ couplings to up-type quarks are flavour diagonal.

In the fermion mass basis, the $Z^{\prime}$ couplings to quarks are then described by
\begin{align}    \label{eqqa}
\mathcal{L}_{Z^{\prime}}  \supset&\; g^{\prime}   Z_{\mu}^{\prime} \Bigl[   \left(    \bar d_i \, \gamma^{\mu} \, P_L \, d_j \, \lambda_{ij}^{dL}    +  \bar d_i \, \gamma^{\mu} \, P_R \, d_j  \, \lambda_{ij}^{dR}            \right) \nonumber \\
&+  \left(    \bar u_i \, \gamma^{\mu} \, P_L \, u_j \, \lambda_{ij}^{uL}    +  \bar u_i \, \gamma^{\mu} \, P_R \, u_j  \, \lambda_{ij}^{uR}            \right)  \Bigr] \,,
  \end{align}
with
\begin{align}
\lambda^{dL} \simeq&\; \begin{pmatrix}
|V_{td}|^2 - \frac{1}{3}     &     V_{ts} V_{td}^*  &    V_{tb}  V_{td}^*   \\
 V_{td} V_{ts}^*    &   |V_{ts}|^2 - \frac{1}{3}  &    V_{tb}  V_{ts}^*  \\
  V_{td} V_{tb}^* &    V_{ts}  V_{tb}^*  &  |V_{tb}|^2 - \frac{1}{3}
\end{pmatrix}  \,, \nonumber \\[0.2cm]
 \lambda^{uL,uR}   \simeq&\;  \lambda^{dR} \simeq \;    \mathrm{diag}(   -1/3, -1/3, +2/3 ) \,.
\end{align}
\section{$\mathbf{B \to K^{(*)} \mu^+ \mu^-}$ anomalies} \label{btok}

In order to have tree-level $Z^{\prime}$ contributions in $b \to s \mu^+ \mu^-$ transitions we extend the model with fermion exotics.    We add to the particle content a vector-like fermion with the following $\mathrm{SU(3)}_C \times \mathrm{SU(2)}_L \times \mathrm{U(1)}_Y \times \mathrm{U(1)}^{\prime}$ quantum numbers
\begin{align}
\psi(1,2,-\frac{1}{2}, X)  \,.
\end{align}
Such vector-like fermion mixes with the $\mathrm{SU(2)}_L$ lepton doublet via the interaction term
\begin{align}   \label{intrr}
\mathcal{L}_Y =  - y_{i} \,  \phi \, \bar \psi \,  P_L\, L_{i} + \mathrm{h.c.} \,,
\end{align}
with $i=\{e,\mu,\tau\}$ a family index.  We assume negligible mixing effects with the electron and the $\tau$-lepton for simplicity.   The $Z^{\prime}$ boson couples to the vector-like fermion as $g^{\prime}  X Z_{\mu}^{\prime}  \bar \psi \gamma^{\mu} \psi $.    Diagonalization of the fermion mass terms gives the following interactions
\begin{align}  \label{eqlab}
\mathcal{L}_{Z^{\prime}} \simeq&\;   g^{\prime }  Z_{\alpha}^{\prime}    \Bigl[   \lambda^{\mu L}  \bar L_{\mu}  \gamma^{\alpha}  P_L    L_{\mu}  +       \lambda^{\psi L}  \bar \psi  \gamma^{\alpha}  P_L    \psi       + X   \bar \psi   \gamma^{\alpha}  P_R      \psi     \nonumber \\
&+    \sqrt{ \lambda^{\mu L} \lambda^{\psi L}     }    \,   \left(  \bar L_{\mu}     \gamma^{\alpha} P_L  \psi  +  \bar \psi     \gamma^{\alpha} P_L  L_{\mu} \right)     \Bigr]
   \,,
\end{align}
with
\begin{align} \label{myleqs}
\lambda^{\mu L}    =&  \frac{X}{ 1+   m_{\psi}^2/(y_{\mu}  \langle \phi \rangle)^2  }   \,, \qquad \lambda^{\psi L}    =  \frac{X}{ 1+   (y_{\mu}  \langle \phi \rangle)^2/m_{\psi}^2  }  \,.
\end{align}
The $Z$ and $W$ couplings are not affected by the fermion mixing given that the vector-like fermion has the same quantum numbers as the $\mathrm{SU(2)}_L$ lepton doublet under the SM-gauge group.  The Dirac mass of the vector-like fermion $m_{\psi}$ is a priori unrelated to the any other mass scale of the theory.   We assume that $m_{\psi}$ is of the order of the $\mathrm{U(1)}^{\prime}$ symmetry breaking scale or lower, $m_{\psi} \lesssim \mathcal{O}(\langle  \phi  \rangle)$.       For a Yukawa coupling $y_{\mu} \sim \mathcal{O}(1)$ one obtains then $ \lambda^{\mu L}, \lambda^{\psi L}   \sim \mathcal{O}(1)$.

The flavour changing transitions $b \to s \mu^+ \mu^-$ are described model-independently in terms of the effective weak Hamiltonian
 \be
   \mathcal{H}_{\rm{eff}}  \supset - \frac{  4 G_F }{\sqrt{2}}   \frac{\alpha}{4 \pi}   V_{ts}^*  V_{tb} \sum_{i=9,10} \left(  C_{i}^{\mu}   Q_{i}^{\mu} + C_{i}^{\prime \mu}   Q_{i}^{\prime \mu}   \right) \,,
 \ee
with $\alpha = e^2/(4 \pi)$ and
\begin{align}
Q_{9}^{\mu} &= (  \bar s \gamma_{\alpha} P_L b  ) (  \bar \mu \gamma^{\alpha} \mu )  \,, \qquad Q_{9}^{\prime \mu} = (  \bar s \gamma_{\alpha}   P_R b  )( \bar \mu \gamma^{\alpha} \mu ) \,, \nonumber \\
Q_{10}^{\mu} &= (  \bar s \gamma_{\alpha} P_L b  ) (  \bar \mu \gamma^{\alpha}  \gamma_5 \mu )  \,, \quad Q_{10}^{\prime \mu} = (  \bar s \gamma_{\alpha}   P_R b  )( \bar \mu \gamma^{\alpha}  \gamma_5 \mu ) \,.
\end{align}
The $Z^{\prime}$ contribution to these Wilson coefficients $C_{i}^{\mu} = C_{i,\rm{SM}}^{\mu} + \delta C_{i}^{\mu}$ is
\begin{align} \label{eq2}
\delta C_{9}^{\mu} &=   - \frac{  \pi    }{   \alpha      }    \frac{     \, g^{\prime \,2}   v^2 \,   \lambda^{\mu L}}{M_{Z^{\prime}}^2}  \,, \quad \delta C_{10}^{\mu}  = - \delta  C_{9}^{\mu} \,.
\end{align}
The chirality-flipped operators $Q_{9,10}^{\prime \mu}$  do not receive new physics contribution in this model.   In the SM $C_{9,\rm{SM}}^{\ell} \simeq - C_{10,\rm{SM}}^{\ell}  \simeq 4.2$.   We neglect $Z$ exchange contributions to $b \to s \mu^+ \mu^-$ since the $Z- Z^{\prime}$ mass mixing is suppressed for large $\tan \beta$ values.      Current $b \to s \ell^+ \ell^-$ data gives the following $3\sigma$ bound on this scenario~\cite{Descotes-Genon:2015uva}:
\begin{align}  \label{wcbound}
\delta C_{9}^{\mu} \in [-1.13, -0.21]  \; \Rightarrow \;   \frac{     \, g^{\prime \,2}   v^2 \,   \lambda^{\mu L}}{M_{Z^{\prime}}^2}   \in [0.5,2.7] \times 10^{-3}  \,.
\end{align}
Similar allowed values for $\delta C_{9}^{\mu}$ are also obtained in~\cite{Altmannshofer:2015sma,Hurth:2016fbr}.\footnote{In contrast, an analysis reduced to the baryonic decays $\Lambda_b \to \Lambda \mu^+ \mu^-$ together with $B_s \to \mu^+ \mu^-$ and inclusive $b \to s \ell^+ \ell^-$ decays shows preference for positive values of $\delta C_{9}^{\mu}$~\cite{Meinel:2016grj}.    }   Note that this bound fixes the sign of $\lambda^{\mu L}$ to be positive and therefore one is left with only two possible $\mathrm{U(1)}^{\prime}$ charge assignments $X=1, 1/2$.

\subsection{Flavour and collider constraints}

Searches for a $Z^{\prime}$ boson at the LHC in the $\mu^+ \mu^-$ decay channel place relevant bounds on the model.    The $Z^{\prime}$ boson will be produced on-shell at the LHC via Drell-Yan processes mainly due to its coupling to the first quark generation.   The $Z^{\prime}$ decay width into a fermion pair $f_i \bar  f_j$ is given by
\begin{align}  \label{eqdecaysZ}
&\Gamma(Z^{\prime} \rightarrow  f_i \bar f_j) \simeq \frac{ N_C M_{Z^{\prime}}   }{  24 \pi } \,  g^{\prime \,2}  \left( |\lambda_{ij}^{f_L}|^2 + |\lambda_{ij}^{f_R}|^2 \right)   \,,
\end{align}
with $N_C = 3 (1)$ for quarks (leptons) and neglecting fermion mass effects.    If the $Z^{\prime}$ decays mainly into SM fermions the branching fraction for $Z^{\prime} \to \mu^+ \mu^-$ is approximately given by
\begin{align}
\mathrm{Br}(Z^{\prime} \to \mu^+ \mu^-) \simeq \frac{    (\lambda^{\mu L})^2 }{   2 ( \lambda^{\mu L} )^2    + 8 } \,.
\end{align}
This would be the case if $m_{\psi} \gtrsim M_{Z^{\prime}}$.     The total width of the $Z^{\prime}$ boson is then given by
\begin{align}
\frac{\Gamma_{Z^{\prime}}}{M_{Z^{\prime}}} \simeq  \frac{g^{\prime \, 2}  \left[ 2   (\lambda^{\mu L})^2  + 8  \right]   }{24 \pi} \,.
\end{align}
For $g^{\prime} \leq 0.6$ and $|\lambda^{\mu L}| \leq 1$ the $Z^{\prime}$ width remains relatively narrow $\Gamma_{Z^{\prime}}/M_{Z^{\prime}} \lesssim 4\%$.    For $m_{\psi}  \lesssim  M_{Z^{\prime}} $, $Z^{\prime}$ decays into a heavy lepton and a muon (or a muon neutrino) become kinematically open while for  $2  m_{\psi}  < M_{Z^{\prime}}$ the $Z^{\prime}$ would decay also into a pair of heavy leptons.  These additional decays would enhance the total $Z^{\prime}$ width.   The $Z^{\prime}$ boson would also couple to a $W^+W^-$ pair due to mixing effects,
\begin{align}
\mathcal{L}_{3} \supset& \;     -  i e \, g_{WWZ^{(\prime)}}  \Bigl[     (  \partial^{\mu} W^{\nu} - \partial^{\nu}  W^{\mu} )   W_{\mu}^{\dag}   Z^{(\prime)}_{\nu}    \nonumber \\
   & -   (  \partial^{\mu} W^{\nu \dag}   - \partial^{\nu} W^{\mu \dag}    )  W_{\mu} Z^{(\prime)}_{\nu}      \nonumber \\
   & +W_{\mu}   W_{\nu}^{\dag}    (  \partial^{\mu}  Z^{(\prime) \nu}    - \partial^{\nu}   Z^{(\prime) \mu}  )   \Bigr] \,,
\end{align}
with $g_{WWZ} \simeq \cot \theta_W$ and $g_{WWZ^{\prime}} = - \xi \cot \theta_W$. The decay rate of the $Z^{\prime}$ into a pair of charged vector bosons is given by
\be\label{eq:zprimetoww}
\Gamma(Z^{\prime}  \rightarrow W^+ W^- ) \simeq \frac{e^2 \cot^2 \theta_W \xi^2  }{   192 \pi } \frac{M_{Z^{\prime}}^5}{M_W^4}  \,.
\ee
Due to the parametrical suppression of $\xi$ in the large $\tan \beta$ limit, see \Cref{emix}, the contribution of this decay channel to the total $Z^{\prime}$ width is negligible.

The $Z^{\prime}$ boson decay into a $Z$ boson and a neutral Higgs is not suppressed by the $Z-Z^{\prime}$ gauge mixing since the scalar doublet $H_1$ carries $\mathrm{U(1)}^{\prime}$ charge.   Assuming decoupling in the scalar sector, the decay channel $Z^{\prime} \to Z h_1$ will be kinematically open, with $h_1$ denoting a SM-like Higgs.  The relevant interaction term is given by
\begin{align}
\mathcal{L} \supset 2 M_Z    g^{\prime}   c_{\beta}^2    Z_{\mu} Z^{\prime \mu}  h_1 \,,
\end{align}
where small corrections due to scalar and gauge boson mixing have been neglected.   The associated decay rate
\begin{align}
\Gamma(Z^{\prime} \to Z h_1) \simeq \frac{g^{\prime 2}  c_{\beta}^4}{  48 \pi }  M_{Z^{\prime}} \,,
\end{align}
receives a strong suppression in the large $\tan \beta$ limit, making its contribution to the total $Z^{\prime}$ width negligible.

We calculate the total $pp \to Z^{\prime} \to \mu^+ \mu^-$ cross-section in the narrow-width approximation $\sigma(pp \to Z^{\prime}) \times \mathrm{Br}(Z^{\prime} \to \mu^+ \mu^-)$ and use the experimental limits obtained by the ATLAS and CMS collaborations at $\sqrt{s} =8$~TeV center-of-mass energy with an integrated luminosity of about $20$~fb$^{-1}$~\cite{Aad:2014cka,Khachatryan:2014fba}.    We also include ATLAS limits on $pp \to Z^{\prime} \to \mu^+ \mu^-$ at $\sqrt{s} =13$~TeV center-of-mass energy with $13.3$~fb$^{-1}$ integrated luminosity~\cite{ATLAS:2016ichep}.   The $Z^{\prime}$ production cross-section at the LHC is estimated using~{\tt MadGraph (MG5\textunderscore aMC\textunderscore 2.4.2)}~\cite{Alwall:2014hca}.    For typical couplings around $g^{\prime} \sim  \lambda^{\mu L}\sim  0.5$ we obtain a lower bound $M_{Z^{\prime}} \gtrsim 3$~TeV.         Current limits on $Z^{\prime} \to t \bar t$ searches set weaker constraints on the model~\cite{CMS:2012rya}.

The $\mathrm{SU(2)}_L$ components of the vector-like fermion $\psi^T = ( N, E )$ are quasi-degenerate.    These states couple to the $W$ and $Z$ bosons like the SM $\mathrm{SU(2)}_L$ lepton doublet and can be pair-produced at the LHC via Drell-Yan processes.  They would decay into bosons ($Z^{(\prime)}, W$ or a Higgs) and charged leptons or neutrinos.  Searches for such states at the LHC are however not very constraining for TeV scale masses~\cite{Falkowski:2013jya,Dermisek:2014qca}.

\begin{figure*}
\centering
\includegraphics[width=7cm]{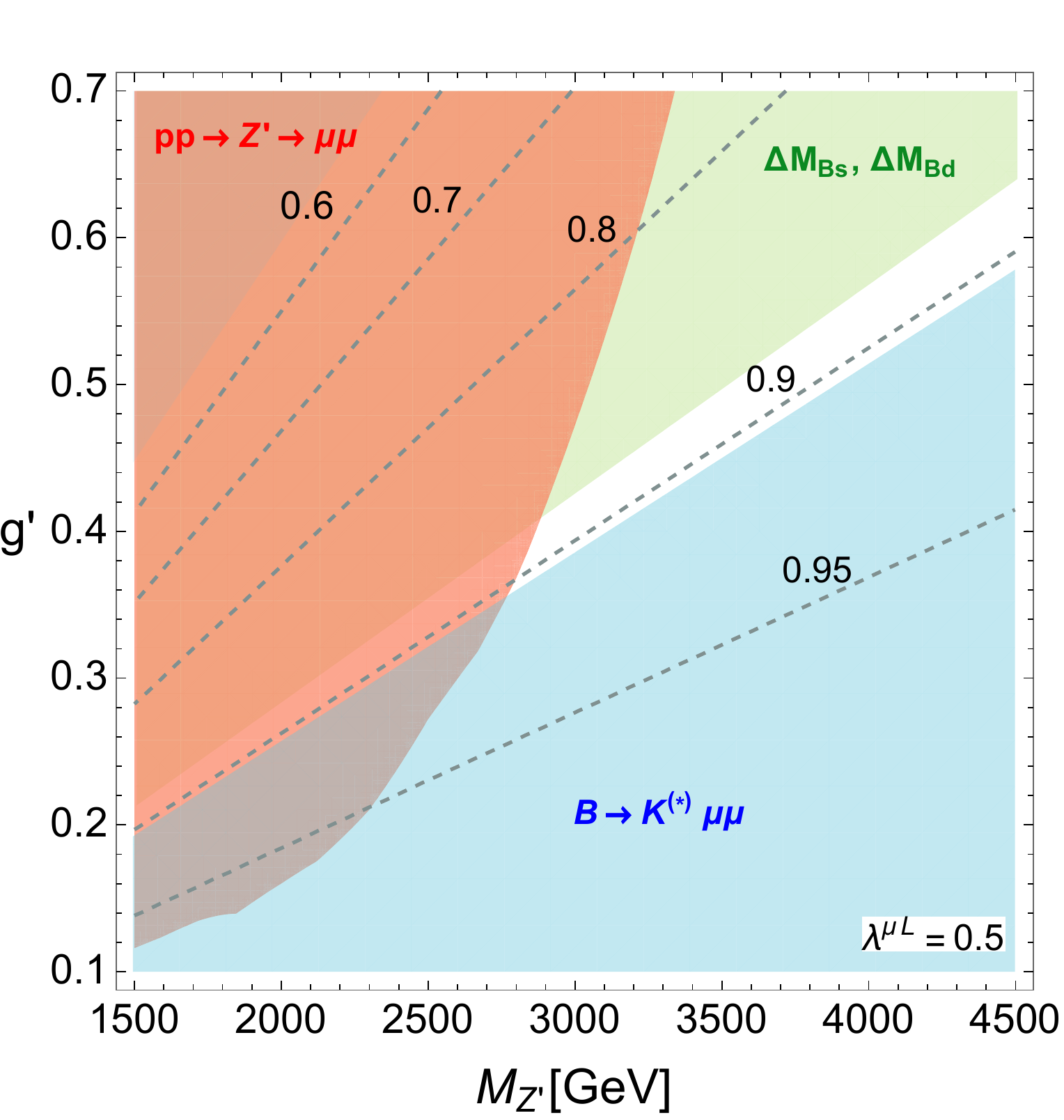}
~~~
\includegraphics[width=7cm]{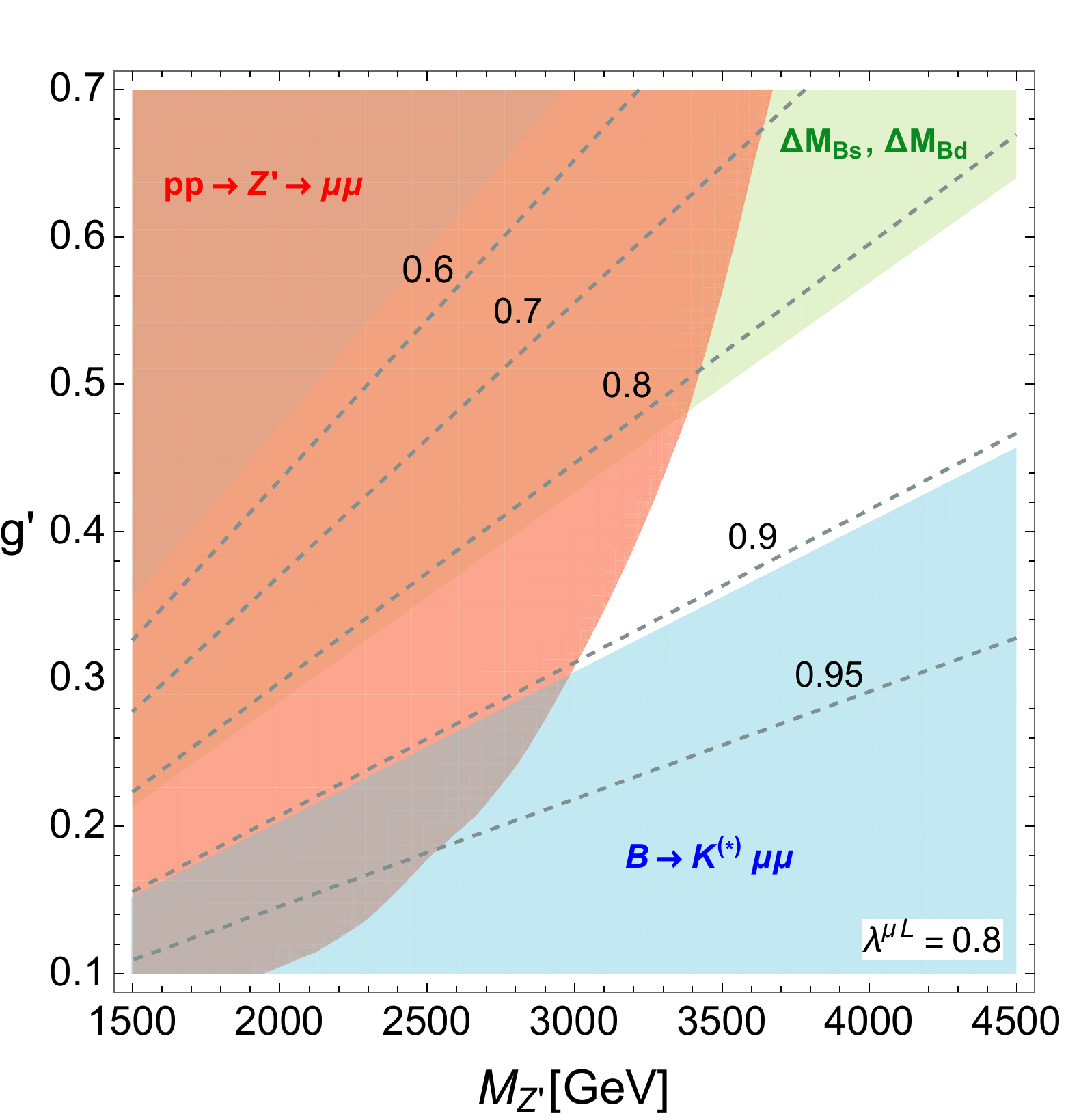}
\caption{\small  \textit{Excluded regions by LHC searches for $pp \to Z^{\prime} \to  \mu^+ \mu^-$ (red), neutral $B$-meson mass differences $\Delta M_{s,d}$ (green) and $B \to K^{(*)} \mu^+ \mu^-$ (blue).    The strength of the $Z^{\prime}$ coupling to muons has been fixed to the benchmark $\lambda^{\mu L} = 0.5$ (left) and $\lambda^{\mu L} = 0.8$ (right).  The white area remains allowed.   Contours of constant $R_K$ are shown as dashed lines. }}
\label{fig:plot1}
\end{figure*}

Another relevant constraint on the model parameter space can be derived from the measured mass differences in the neutral $B_q$-meson system ($q=d,s$).  The $Z^{\prime}$ gives a tree-level contribution to the mass difference $\Delta M_q = \Delta M_q^{\rm{SM}} + \Delta M_q^{Z^{\prime}}$:
\begin{align}
\Delta M_q^{Z^{\prime}} =   \frac{1}{3}  M_{B_q}  f_{B_q}^2  \hat B_{Bq}^{(1)} \, \eta_{11}   \frac{    g^{\prime \,2}   | \lambda_{qb}^{dL} |^2    }{M_{Z^{\prime}}^2}     \,,
\end{align}
with $\eta_{11}(\sim1~\text{TeV}) \simeq 0.8$ accounting for renormalization group effects and $f_{B_q}^2  \hat B_{Bq}^{(1)}$ representing a hadronic quantity that must be determined using non-perturbative methods~\cite{Buras:2012dp,Buras:2012jb,Lenz:2010gu}.  We use CKM inputs determined from tree-level processes and hadronic inputs from Lattice QCD~\cite{CKMgroup,1602.03560}.     Note that the $Z^{\prime}$ contribution always enhances $\Delta M_q$ with respect to the SM. Due to the flavour structure of the model we have
\begin{align}
\frac{\Delta M_s^{Z^{\prime}} }{\Delta M_d^{Z^{\prime}} } =   \frac{ \Delta M_s^{\rm{SM}}   }{\Delta M_d^{\rm{SM}}  }   =   \frac{  M_{B_s}  f_{B_s}^2  \hat B_{Bs}^{(1)}   }{   M_{B_d}  f_{B_d}^2  \hat B_{Bd}^{(1)} }   \left|  \frac{V_{ts}}{V_{td}} \right|^2  = 31.2 \pm 1.8 \,.
\end{align}
Using the current experimental values for $\Delta M_{s,d}$~\cite{1412.7515}, we obtain the following bound at the $3\sigma$ level
\begin{align}
\frac{ g^{\prime} v }{M_{Z^{\prime}}} < 0.035  \,.
\end{align}

The constraints from $B \to K^{(*)} \mu^+ \mu^-$, neutral $B$-meson mass differences and LHC searches for a $Z^{\prime}$ boson in the $\mu^+ \mu^-$ channel are illustrated in Fig.~\ref{fig:plot1} for the benchmark values $\lambda^{\mu L } = 0.5, 0.8$.   With the bounds considered we find the lower bound $\lambda^{\mu L } \gtrsim 0.41$.   Contours of constant $R_K$ are also shown in the plot as dashed lines.      The ratio $R_K$ is approximately given in the model by~\cite{Hiller:2014ula}
\begin{align}
R_K \simeq 1 +    \frac{  2 \delta C_{9}^{\mu}   }{  C_{9,\rm{SM}}^{\mu} }  +  \left( \frac{\delta C_{9}^{\mu}}{C_{9,\rm{SM}}^{\mu}} \right)^2 \,.
\end{align}
For $\lambda_{\mu L } = 0.5$ the allowed values for this ratio are around $R_K \simeq 0.9$ as shown in Fig.~\ref{fig:plot1}.  This is compatible within $2 \sigma$ with the LHCb measurement $R_K^{\mbox{\scriptsize{LHCb}}} = 0.745^{+0.090}_{-0.074}\pm0.036$~\cite{Aaij:2014ora}.   Taking into account the flavour and LHC constraints discussed we find $R_K \in [0.8,0.9]$ when considering $\lambda_{\mu L } \lesssim 1$.        Since the flavour changing $Z^{\prime}$ couplings to down-type quarks are left-handed, a test of the model would be the measurement of $R_{K^*}$~\cite{Hiller:2014ula}.   In our model we expect $R_{K^*} \simeq R_K$.

New physics also enters at tree-level in the decays $b \to s \nu  \bar \nu$.    Due to $\mathrm{SU(2)}_L$ gauge symmetry, the leading new physics contribution to the effective weak Hamiltonian is related to that in $b \to s \mu^+ \mu^-$.  Namely, the $Z^{\prime}$ exchange generates the four fermion operator
\begin{align}
   \mathcal{H}_{\rm{eff}}  \supset - \frac{  4 G_F }{\sqrt{2}}   \frac{\alpha}{4 \pi}   V_{ts}^*  V_{tb}   C_{L}^{sb, \nu} Q_{L}^{sb, \nu}     + \mathrm{h.c.} \,,
\end{align}
with $Q_{L}^{sb, \nu}   = [  \bar s \gamma_{\alpha}   P_L b  ][  \bar \nu_{\mu}   \gamma^{\alpha}   (1 - \gamma_5) \nu_{\mu} ] $ and
\be C_{L}^{sb, \nu} = C_{L}^{sb, \nu}\bigr|_{\rm{SM}} +  \delta C_{9}^{\mu} \,.
\ee
The SM contribution $C_{L}^{sb,\nu}\bigr|_{\rm{SM}}\simeq -6.4$ has been detailed in~\cite{Buras:2014fpa} and $\delta C_{9}^{\mu}$ was given in \eqref{eq2}.  Taking into account the non-universal character of the new physics~\cite{Buras:2014fpa}, the exclusive decays $B \to K^{(*)} \nu \bar \nu$  are described by
\begin{align}
R_{K^{(*)}}  = \frac{  \mathrm{Br}(B \to K^{(*)} \nu \bar \nu ) }{ \mathrm{Br}(B \to K^{(*)} \nu \bar \nu )\bigr|_{\rm{SM}}}  = \frac{1}{3}  (2 +  \epsilon_{\mu}^2)  \,,
\end{align}
with
\begin{align}
\epsilon_{\mu}  =    \frac{\sqrt{  | C_{L}^{sb,\nu}\bigr|_{\rm{SM}} + \delta C_{9}^{\mu}    |^2  }  }{|C_{L}^{sb,\nu}\bigr|_{\rm{SM}} |} \,.
\end{align}
The ratios $R_{K^{(*)}}$ receive an enhancement with respect to the SM since the negative sign of $\delta C_{9}^{\mu}$ preferred by experiments causes a constructive interference between the new physics part and the SM.   Flavour and LHC constraints however restrict this enhancement to be very small, $R_{K^{(*)}} \lesssim 1.05$, well below the current experimental limits $R_{K^{(*)}}  \lesssim 4.4$~\cite{Lees:2013kla,Lutz:2013ftz}.

Finally, we note that one-loop contributions to the muon anomalous magnetic moment with an internal vector-like fermion $(\psi)$ are negligible for $m_{\psi} \sim \langle \phi \rangle$ and $y_{\mu} \sim \mathcal{O}(1)$; given that the necessary chirality flip only occurs through a mass insertion in the external lepton lines.

\subsection{$B_{s,d} \to \mu^+ \mu^-$}

The decays $B_{s,d} \to \mu^+ \mu^-$ constitute a clean mode to test for small new physics effects.   These decays are sensitive to tree-level neutral Higgs exchange contributions encoded in scalar and pseudo-scalar operators
 \be
   \mathcal{H}_{\rm{eff}}  \supset - \frac{  4 G_F }{\sqrt{2}}   \frac{\alpha}{4 \pi}   V_{ts}^*  V_{tb} \sum_{i=S,P} \left(   C_{i}^{\mu}   Q_{i}^{\mu} + C_{i}^{\prime \mu}   Q_{i}^{\prime \mu}   \right) \,,
 \ee
with
\begin{align}
Q_{S}^{\mu} &=  m_b (  \bar s  P_R b  ) (  \bar \mu   \mu )  \,, \qquad Q_{S}^{\prime \mu} =   m_b (  \bar s    P_L b  )( \bar \mu  \mu ) \,, \nonumber \\
Q_{P}^{\mu} &=  m_b (  \bar s  P_R b  ) (  \bar \mu   \gamma_5 \mu )  \,, \quad Q_{P}^{\prime \mu} =  m_b (  \bar s    P_L b  )( \bar \mu  \gamma_5 \mu ) \,.
\end{align}
Tree-level new physics contributions to $B_{s,d} \to \mu^+ \mu^-$ in our model are described by~\cite{Hiller:2014yaa}
\begin{align}
\frac{  \mathrm{Br}(  \bar B_q \to \mu^+ \mu^- )  }{\mathrm{Br}(  \bar B_q \to \mu^+ \mu^- )_{\rm{SM}} } \simeq  |     1 - 0.24  \delta C_{10}^{\mu}  - \kappa C_{P}^{\mu}    |^2  +  |  \kappa  C_{S}^{\mu}    |^2 \,,
\end{align}
with $\kappa \simeq 7.7 m_{b}$.     The scalar sector of the model is assumed to be in a decoupling regime, with a light SM-like Higgs boson $h_1$ to be identified with the $125$~GeV boson and additional heavy scalars at the TeV scale.   Among the heavy scalars we have: two CP-even Higgs bosons $h_{2,3}$, a CP-odd Higgs $A$ and a charged Higgs $H^{\pm}$.    We denote by $h_2$ the scalar coming mainly from the Higgs doublets while $h_3$ is basically the real part of the scalar field $\phi$.   Details of the scalar sector are provided in Appendix~\ref{aop:sec}.

The dominant tree-level contributions to $B_{s,d} \to \mu^+ \mu^-$ due to Higgs exchange arise from $h_2$ and $A$, their Yukawa couplings to down-type quarks and leptons read
\begin{align}
-\mathcal{L}_Y &\supset   \frac{h_2}{v} \left(  \bar d_L \,N_d  \, d_R     + \frac{1}{t_{\beta}}    \bar \ell_L \, M_{\ell}^{\rm{diag}} \, \ell_R    \right)   \nonumber \\
&+ i \frac{A}{v} \left(   \bar d_L \, N_d \, d_R  +    \frac{1}{t_{\beta}}    \bar \ell_L \, M_{\ell}^{\rm{diag}} \, \ell_R     \right)  + \mathrm{h.c.} \,.
 \end{align}
Here $N_d$ is expressed in the fermion mass basis
\begin{align}
N_d &= \,   \frac{1}{t_{\beta}} M_d^{\rm{diag}}  - \frac{v}{ \sqrt{2} }  \frac{1}{s_{\beta}}  ( V_{dL}^{\dag}  \Gamma_1  V_{dR}  ) \,, \nonumber \\
&\simeq   \frac{1}{t_{\beta}} M_d^{\rm{diag}}   -  \frac{m_b}{  s_{\beta}  c_{\beta} }  \begin{pmatrix}
0 & 0 & - V_{td}^*  V_{tb}\\
0 & 0 & - V_{ts}^*  V_{tb} \\
0 & 0 &   1 - |V_{tb}|^2
\end{pmatrix} \,.
\end{align}
The flavour diagonal couplings to quarks and leptons show the typical Type-I structure of two-Higgs-doublet models and are suppressed for large $\tan \beta$.  The scalar and pseudo-scalar Wilson coefficients are given by
\begin{align}
C_{S}^{\mu}   = -  C_{P}^{\mu} =    \frac{2  \pi   }{\alpha  s_{\beta}^2   }   \frac{m_{\mu}}{\bar M_h^2}   \,.
\end{align}
The Higgs bosons $h_2$ and $A$ are quasi-degenerate so we denote $\bar M_h^2 \equiv M_{h_2}^2 \simeq M_A^2$ (up to corrections of $\mathcal{O}(v^2)$).   Scalar contributions to the chirality-flipped operators $Q_{S, P}^{\prime \mu}$ are suppressed by a factor $m_s/m_b$ and are neglected.  To a good approximation the ratio $\mathrm{Br}(\bar B_d \to \mu^+ \mu^-)/\mathrm{Br}(\bar B_s \to \mu^+ \mu^-)$ remains unchanged with respect to the SM in this model.

We find that scalar contributions to $B_{s,d} \to \mu^+ \mu^-$ are negligible in the model for $\bar M_h  \sim \mathcal{O}(\text{TeV})$.  The $Z^{\prime}$ contribution on the other hand causes a significant suppression of $\mathrm{Br}(B_{s,d} \to \mu^+ \mu^-)$ with respect to the SM.    Taking into account the flavour and LHC constraints discussed previously we find $\mathrm{Br}(B_{s,d} \to \mu^+ \mu^-)$ to be between $10\%$ and $20\%$ below the SM prediction, depending on the value of $\lambda_{\mu L}$.   Such values for  $\mathrm{Br}(B_{d} \to \mu^+ \mu^-)$  are in slight tension with the measured branching fraction~\cite{CMS:2014xfa}
\begin{align}
 \mathrm{Br}(B_{d} \to \mu^+ \mu^-)_{\rm{LHCb+CMS}} = (3.9^{+1.6}_{-1.4}) \times 10^{-10} \,,
\end{align}
which lies at face value above the SM prediction  $\mathrm{Br}(B_{d} \to \mu^+ \mu^-)_{\rm{SM}} = (1.06 \pm 0.09 ) \times 10^{-10}$~\cite{Bobeth:2013uxa}.  Current measurements of the $B_s \to \mu^+ \mu^-$ decay~\cite{CMS:2014xfa,Aaboud:2016ire}
\begin{align}
\mathrm{Br}(B_{s} \to \mu^+ \mu^-)_{\rm{LHCb+CMS}} &= (2.8^{+0.7}_{-0.6}) \times 10^{-9} \,, \nonumber \\
\mathrm{Br}(B_{s} \to \mu^+ \mu^-)_{\rm{ATLAS}} &= (0.9^{+1.1}_{-0.8}) \times 10^{-9} \,,
\end{align}
are compatible with the SM prediction $\mathrm{Br}(B_{s} \to \mu^+ \mu^-)_{\rm{SM}} =  (3.66 \pm 0.23) \times 10^{-9}$~\cite{Bobeth:2013uxa}.

\section{Dark Matter}  \label{sec:dmd}

The gauged family baryon number can also play an additional role in connection to DM.   Indeed, a fermion carrying only $\mathrm{U(1)}^{\prime}$ quantum numbers would be a natural DM candidate.      Consider a vector-like fermion with $\mathrm{SU(3)}_C \times \mathrm{SU(2)}_L \times \mathrm{U(1)}_Y \times \mathrm{U(1)}^{\prime}$ quantum numbers
\begin{align}
\chi(1,1,0,B_{\chi}) \,.
\end{align}
The only renormalizable terms involving $\chi$ in the Lagrangian are
\begin{align}
\mathcal{L}_{\chi}  =    \left[   \bar \chi \left(    i   \slashed{D}  - m_{\chi}   \right)   \chi \right]    \,.
\end{align}
The Dirac mass $m_{\chi}$ is a priori unrelated to other mass scales of the theory but we assume $m_{\chi} \lesssim \mathcal{O}(v_{\phi})$.  The covariant derivative in this case reads $D_{\mu} = \partial_{\mu} - i g^{\prime}  B_{\chi}  \hat Z^{\prime}_{\mu}$.  After spontaneous symmetry breaking, a stabilizing discrete $\mathcal{Z}_2$ symmetry remains in the Lagrangian under which $\chi$ is odd and all the other fields are even.   The fermion $\chi$ is therefore stable due to the underlying $\mathrm{U(1)}^{\prime}$ local symmetry.

Generically, we consider this model to be an effective theory valid up to a cut-off scale $\Lambda \gg v_{\phi}$ where new dynamics is supposed to appear.  At low-energies, $E \ll \Lambda$, the effects of the high-scale dynamics is parametrized by a tower of effective operators of increasing canonical dimension.     Majorana neutrino masses, for instance, would arise from the dimension five Weinberg operator $(\bar L_L^{c}   \tilde H_2^{*}) (   \tilde H_2^{\dag}   L_L )$ after spontaneous symmetry breaking~\cite{Weinberg:1979sa}.

Some of the effective operators could break the $\mathcal{Z}_2$ symmetry explicitly, upsetting the stability of $\chi$.       Our DM candidate is neutral under the SM gauge group and carries family baryon number, so that it could balance the family baryon number carried by a $\mathrm{SU(3)}_C \times \mathrm{SU(2)}_L \times \mathrm{U(1)}_Y$ gauge invariant operator with field content $(u_R  d_R  d_R)$.    Operators of the type $\chi ( u_R  d_R  d_R)^m$ could then be invariant under the full gauge group, introducing explicit breaking terms of the $\mathcal{Z}_2$ symmetry and potentially destabilizing the DM candidate.   These dangerous effective operators would be forbidden by the gauge symmetry as long as $B_{\chi}$ is not integer.\footnote{Note that baryon number is not protected by the family symmetry considered $\mathrm{U(1)}^{\prime} =  B_1 + B_2 -2 B_3$.   Dimension six operators violating baryon number are allowed by the full gauge group and will contribute to baryon number violating processes~\cite{Weinberg:1979sa}.}

Taking into account the $Z-Z^{\prime}$ mass mixing, the interactions of $\chi$ read
\begin{align}  \label{eqmixf}
\mathcal{L}_{\chi}  \supset  g^{\prime}  B_{\chi}    \left(   Z^{\prime}_{\mu}   + \xi    Z_{\mu} \right)  \bar \chi \gamma^{\mu}   \chi    \,, \quad   \xi  \approx -\frac{e g^{\prime}}{2 c_W s_W }  \frac{   c_{\beta}^2 v^2     }{   M_{Z^{\prime}}^2  }     \,.
\end{align}
The massive neutral vector bosons of the theory would act as a vector portal between the dark sector and the SM particles.     Gauge mixing effects are suppressed in the large $\tan \beta$ limit and are going to be neglected in the following unless explicitly stated.    The DM candidate can in principle annihilate through $Z^{\prime}$ exchange into SM fermion pairs and satisfy the current observed value of the DM relic density.

\subsection{Relic density}
%

\begin{figure}
\centering
\includegraphics[width=8.5cm]{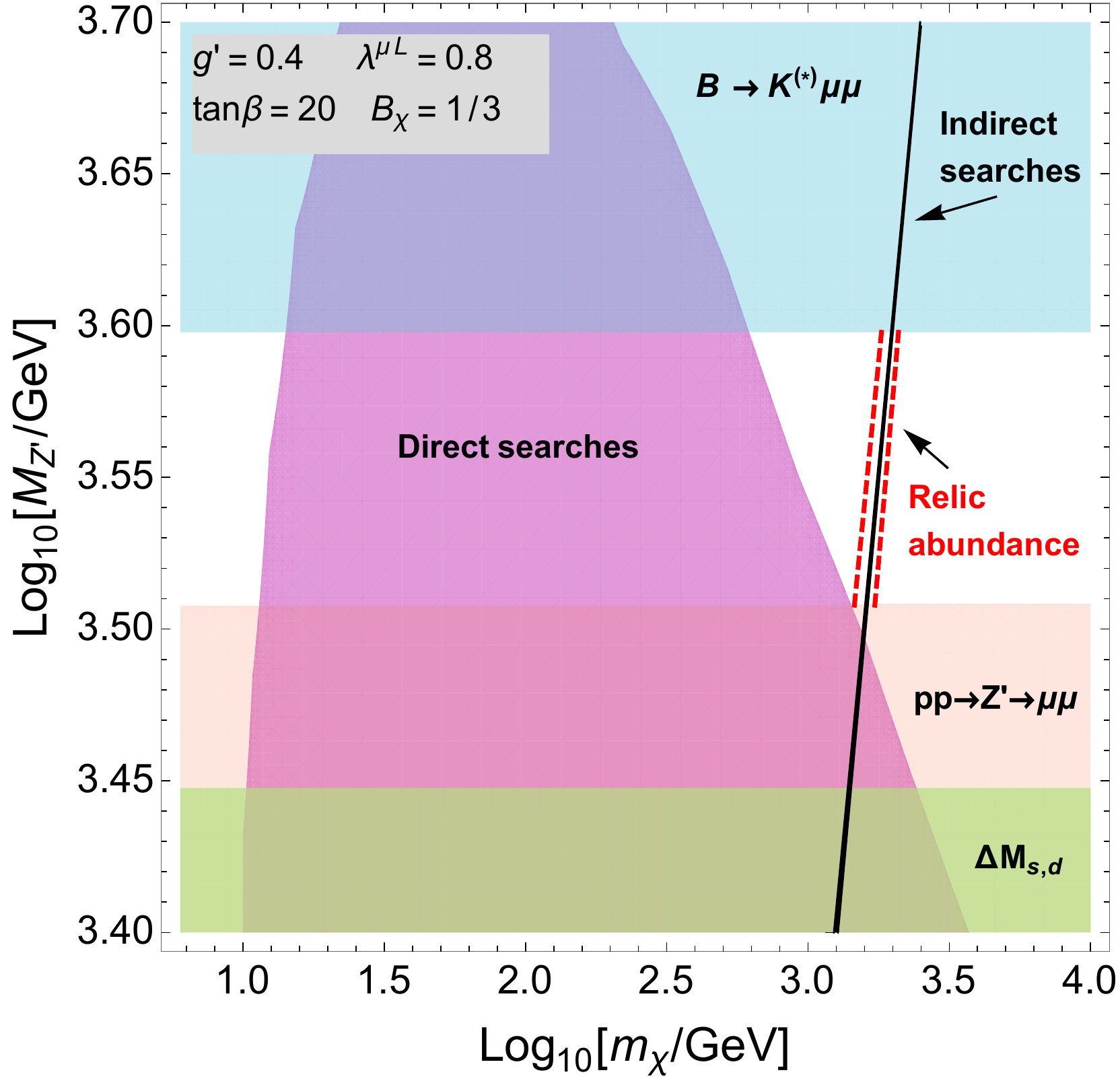}
\caption{\small  \textit{Excluded regions by $\Delta M_{s,d}$ (green), $pp \to Z^{\prime} \to \mu^+ \mu^-$ (red), $b \to s \ell^+ \ell^-$ (blue), direct DM searches (purple) and indirect DM searches (black line).  Values accommodating the observed DM relic abundance appear as two dashed red lines.    }}
\label{fig:plot2}
\end{figure}

For weakly interacting massive particle (WIMP) DM candidates
the predicted relic density comes out in general close to the measured value.  In our case, due to various constraints discussed in previous sections, the mediator,
i.e., the $Z'$ gauge boson, is much heavier than the weak interaction gauge bosons ($M_{Z^{\prime}} \gtrsim 3~\rm{TeV}$).
Hence we cannot achieve a large enough annihilation cross section with weak scale DM candidates.
In this work, we consider DM annihilating via a narrow Breit-Wigner resonance around threshold.
We rely on the Wigner enhancement effect to obtain the right size of the annihilation cross section according to the observed DM relic abundance.
Thus, in order to accommodate the observed DM relic abundance in the model, we would need a TeV scale DM candidate $\chi$ with $m_\chi \sim M_{Z'}/2$.

For convenience, we write the interactions of $Z'$ to $\chi$ and the SM fermions in the form
\begin{equation}
\mathcal{L}_{Z'}\sim g' B_\chi Z_{\mu}^{\prime}\,\bar{\chi}\gamma^{\mu}\chi+g'Z_{\mu}^{\prime}\Big[\bar{f}\gamma^{\mu}(V_{f}+A_{f}\gamma_{5})f\Big]\,.
\label{Zpintera}
\end{equation}
Explicit values for the vector and axial-vector couplings $V_f,A_f$ can be obtained from \Cref{eqqa},
\begin{align}
V_d &=  \left(  -1/3, -1/3, 2/3  \right) \,, \qquad    A_d =  0 \,, \nonumber \\
V_u &=   \left(  -1/3, -1/3, 2/3  \right)    \,, \qquad    A_u =  0\,, \nonumber \\
V_{\ell} &=(0,    \lambda^{\mu L}/2   , 0)  \,, \qquad \qquad    ~~    A_{\ell} =  (0,  -    \lambda^{\mu L}/2 ,0 ) \,, \nonumber \\
V_{\nu} &= (0,     \lambda^{\mu L}/2 , 0)  \,, \qquad \qquad   \, ~ A_{\nu} = (0,  -   \lambda^{\mu L}/2, 0 ) \,.
\end{align}
Here we have neglected small-off diagonal couplings to down-type quarks and the parameter $\lambda^{\mu L}$ was defined in~\Cref{myleqs}.    The total decay width of $Z'$ is then given by
\begin{align}
\Gamma_{Z'}& =  \sum_{f\in \rm{SM}} \frac{N_{C}M_{Z'}}{12\pi}\sqrt{1-\frac{4m_{f}^{2}}{M_{Z'}^{2}}}   \nonumber  \\
&\times \left[(g'V_{f})^{2}\left(1+\frac{2m_{f}^{2}}{M_{Z'}^{2}}\right)+(g'A_{f})^{2}\left(1-\frac{4m_{f}^{2}}{M_{Z'}^{2}}\right)\right]     \nonumber \\
&+\frac{ (g^{\prime} B_{\chi})^2   M_{Z^{\prime} } }{   12 \pi }  \sqrt{  1 - \frac{4 m_{\chi}^2}{  M_{Z^{\prime}}^2 } }     \left(1  + \frac{2 m_{\chi}^2}{M_{Z^{\prime}}^2} \right) \,,
\end{align}
where we have assumed that the $Z^{\prime}$ boson does not decay into final states involving the vector-like fermion $\psi$ ($m_{\psi} \gtrsim M_{Z^{\prime}}$).

We note that the axial coupling of $Z'$ to DM is zero and thus
the total DM annihilation cross section is given by
\begin{align}
\sigma   v & =\sum_{f}\frac{N_{C}\,(g' B_\chi )^{2}(s+2m_{\chi}^{2})}{  6 \pi  s \left[(s-M_{Z'}^{2})^{2}+M_{Z'}^{2}\Gamma_{Z'}^{2}\right]} \sqrt{1-\frac{4m_{f}^{2}}{s}}\nonumber \\
 & \qquad\times\left[(g'V_{f})^{2}(s+2m_{f}^{2})+(g' A_{f})^{2}(s-4m_{f}^{2})\right]\,,
\label{AnnCS}
\end{align}
where we sum over all of the SM final states and the relative velocity is $v = 2 (1 - 4 m_{\chi}^2/s)^{1/2}$.

As the DM mass approaches $M_{Z'}/2$,
the denominator in the annihilation cross section \Cref{AnnCS}
becomes smaller,
reaching its minimum value when $m_\chi = M_{Z'}/2$.
Thus, when DM annihilates near the pole,
the total DM annihilation cross section receives a significant enhancement.
This effect can give a large boost of $\langle \sigma v \rangle$,
and is known as the Wigner enhancement of DM annihilation~\cite{Gondolo:1990dk,Griest:1990kh,Nath:1992ty,Ibe:2008ye}.

The thermal relic abundance of DM $\chi$ is governed by
the Boltzmann equation of DM number density
\begin{equation}
\frac{{\rm d}Y}{{\rm d}x}=-\sqrt{\frac{\pi}{45G}}\frac{\sqrt{g_{*}}m_{\chi}}{x^{2}}\langle\sigma v\rangle(Y^{2}-Y_{EQ}^{2})\,,
\label{BoltzEq}
\end{equation}
where $G=1/M^2_{\rm Pl}$ is the  gravitational constant, $x\equiv m_{\chi}/T$, $Y\equiv n_{\chi}/\hat s$, the ratio of DM number density over the entropy density, which gives rise to
the number of DM particles per comoving volume, $g_{*}$ is
the total number of effective degrees of freedom.
In our case, TeV scale DM particle freezes out at around $50~\rm{GeV}$,
and thus the corresponding $g_* = 96.25$. The equilibrium value of DM particle
number per comoving volume is given by
\begin{equation}
Y_{EQ}=\frac{n_{EQ}}{\hat s}=\frac{45}{2\pi^{4}}\sqrt{\frac{\pi}{8}}\frac{g_{\chi}}{g_{*}}x^{3/2}{\rm e}^{-x}\,,
\label{YEQ}
\end{equation}
where $g_{\chi}=2$ is the degree of freedom of DM particle $\chi$.
The thermal averaged DM annihilation cross section times the relative velocity is given by~\cite{Gondolo:1990dk}
\begin{equation}
\langle\sigma v\rangle=\frac{{\displaystyle \int}_{4m_{\chi}^{2}}^{\infty}{\rm d}s\,s\sqrt{s-4m_{\chi}^{2}}K_{1}(\sqrt{s}/T)\sigma v}{16Tm_{\chi}^{4}K_{2}^{2}(x)}\,,
\label{sigmav1}
\end{equation}
where $K_{\alpha}$ are the modified Bessel function of order $\alpha$.\footnote{
In~\cite{Griest:1990kh} and many works thereafter,
the Boltzmann distribution was used to calculate thermal
averaged cross section times relative velocity
\begin{equation}
\langle\sigma v\rangle=\frac{x^{3/2}}{2\sqrt{\pi}}\int_{0}^{\infty}{\rm d}v\,v^{2}(\sigma v){\rm e}^{-xv^{2}/4}\,,\nonumber
\end{equation}
in whose approximation a non-relativistic regime was assumed.
In our computation, we find that
the DM relic density obtained by using the above formula can differ by $\sim50\%$.
}

When $Y$ ceases to deviate from $Y_{EQ}$, DM starts to
freeze out. The freeze-out value $x_{f}\equiv m_{\chi}/T_{f}$ is
defined at the time when $Y-Y_{EQ}$ becomes of the same order than $Y_{EQ}$,
and is usually defined by the criterion: $Y-Y_{EQ}=cY_{EQ}(x_{f})$
where $c$ is an $\mathcal{O}(1)$ number that has a small numerical impact in determining the freeze-out value $x_f$. We obtain $x_{f}$ by solving the
Boltzmann equation~\Cref{BoltzEq} iteratively.

Considering the benchmark model with
$g'=0.4$, $\lambda^{\mu L}=0.8$, $\tan \beta = 20$, $B_\chi=1/3$,
and using \Cref{BoltzEq,YEQ,sigmav1}, we obtain $x_f \approx 30$.   The efficiency of the post freeze-out annihilation is expressed by the integral
\begin{equation}
J(x_{f})=\int_{x_{f}}^{\infty}\frac{\langle\sigma v\rangle}{x^{2}}{\rm d}x\,.
\end{equation}
The current relic density of DM is given by
\begin{equation}
\Omega h^{2}\approx \frac{2\times 1.07\times10^{9}\,{\rm GeV}^{-1}}{\sqrt{g_{*}}M_{{\rm Pl}}J(x_{f})}\,,
\end{equation}
where the Planck mass $M_{{\rm Pl}}=1.22\times10^{19}\,{\rm GeV}$, and the additional factor of 2 counts the contribution of both the Dirac DM particle and its anti-particle.

In~\Cref{fig:plot2}, we show the values of the $Z^{\prime}$ and DM mass accommodating the observed DM relic abundance for the benchmark point considered.      As observed earlier, the model can accommodate the DM relic abundance when the DM mass is near the pole $2 m_{\chi} \sim M_{Z^{\prime}}$ due to the Wigner enhancement effect.

\subsection{Indirect Dark Matter detection}

DM annihilations can also be probed by astronomical observations. Particularly interesting is the associated emission of photons (gamma rays). The phenomenology of such emission is trivial since these propagate in straight lines. The gamma ray flux per unit photon energy can be obtained by
\begin{equation}
\label{eq:idmd}
\frac{\dif\Phi_\gamma}{\dif E}=\frac1{16\pi m_\chi^2}J_{\rm target}\times\sigma_{\rm ann}v\sum_c\br_c\frac{\dif N_\gamma}{\dif E}\ .
\end{equation}

This is a rather useful formula where the macroscopic and the microscopic physics are factorized out. On the one hand, the $J$ factor
\begin{equation}
J=\int\dif\Omega\int_{\rm l.~o.~s.}\!\!\!\!\!\dif l\ \rho_\chi^2
\end{equation}
quantifies the intrinsic usefulness of the observed region as target for indirect DM detection: l.~o.~s. stands for line of sight. On the other hand, the signal is also proportional to the ``ability'' for the DM to annihilate into photons. Quantitatively this is given by the cross section times the photon yield per annihilation channel (labelled with the letter $c$ in~\Cref{eq:idmd}).

Due to the simplicity of \Cref{eq:idmd} and the enormous progress in gamma-ray astronomy in recent years most of the studies in indirect DM detection are concerned with this messenger. In particular, dwarf spheroidal (dSph) galaxy observations by the Fermi Gamma-ray Space Telescope provide the up-to-date most stringent constraints on the annihilation cross section of DM in the sub-TeV mass region~\cite{Ackermann:2015zua,Ahnen:2016qkx}. Complementarily, observations by the HESS Telescope of the innermost parts of our Galaxy provide the strongest limits in the trans-TeV mass region~\cite{::2016jja}.

We use therefore the limits reported in~\cite{Ackermann:2015zua,Ahnen:2016qkx,::2016jja} in order to constrain the parameters of the model presented in this note. In doing this we only need the $s$-channel $\bar\chi\chi$ annihilation rates at the center of mass energy $\sqrt s=2m_\chi$.   Relevant bounds on our model are derived from DM annihilation into fermion-antifermion channels, see~\Cref{AnnCS}.  Note that $\bar\chi\chi\to W W$ annihilations are suppressed in our model by the hard dependence of the $\xi$ parameter while the $\bar\chi\chi\to Z'h$ annihilation is kinematically suppressed close to the $m_\chi=M_{Z'}/2$ resonance. Off this resonance DM annihilation rates are too weak to be observed by the Fermi and HESS telescopes.

The strong limits that HESS puts on the $\bar\chi\chi\to t \bar{t}$ and $b \bar b$ annihilation channels can be used to obtain exclusion regions. Notice that since the U(1)$'$ charges of third generation quarks are twice as large as the ones of the other two generations, the annihilation cross section into these two channels are by far the largest.  In particular $\chi\chi\to b\bar b$ annihilations are such that $A_f=0$, $V_f=2/3$ and for large $Z'$ masses near the resonance
\be
\sigma v\sim 10^{-23}~{\rm cm}^3/{\rm s}\left(\frac{\rm TeV}{M_{Z'}}\right)^2 \,,
\ee
independently of the couplings $g'$ and $\lambda^{\mu L}$.

In Fig.~\ref{fig:plot2} indirect detection constraints on the model are also included. The models considered in the figure are better constrained by the HESS observation \cite{::2016jja}.    As evident in the figure, our indirect detection constraint is very narrow and quite sensitive to unaccounted-for effects. These include the radiative processes and halo velocity dispersion effects. However, at the quantitative level, these effects are negligible when compared with the uncertainties of the Galactic inner halo observed by the HESS Telescope.

Concretely, the indirect detection limits that result from considering $J$ factors ten times larger than the ones considered in~\cite{::2016jja} will exclude the models with the correct relic DM density. The latter are defined by the red dashed lines in Fig.~\ref{fig:plot2}. On the other hand, unrealistically shallower profiles ($J$ factors three orders of magnitude smaller) than the ones considered in~\cite{::2016jja} would then be necessary for the DM in this model to be undetectable by indirect detection experiments.

\subsection{Direct detection of Dark Matter}

Particles from the Milky Way DM halo can be detected in principle in terrestrial experiments if these interact with the SM.      Direct detection experiments search for WIMP scattering off a target nucleus by identifying the associated nuclear recoil in a extremely low background environment.   Typical recoil energies expected in these experiments are in the $(1-100)$~keV range~\cite{Undagoitia:2015gya}.

At tree-level, the elastic scattering between a WIMP and a nucleus will be mediated in our model by the $t$-channel exchange of a massive neutral vector boson with vectorial couplings to the DM candidate and to quarks (up to gauge mixing effects).

In order to compute the direct detection rates we match our model to an effective theory at the nuclear energy scale.     Running effects from the $\mathrm{U(1)}^{\prime}$ symmetry breaking scale ($\sim$~TeV) down to the nuclear scale ($\sim$~GeV) have a small numerical impact in this model.\footnote{We evaluate these effects following~\cite{D'Eramo:2016atc}, using the code {\tt runDM}~\cite{runDM}.}      At the nuclear energy scale our model generates dominant contributions to the standard spin-independent interaction to the nucleons ($N= p,n$)
\begin{align}
\mathcal{L}_{\chi N} \simeq    \frac{  g^{\prime \, 2}   B_{\chi}    }{M_{Z^{\prime}}^2} \,    ( \bar \chi \gamma_{\mu} \chi ) ( \bar N \gamma^{\mu} N  ) \,.
\end{align}
The latter is subject to a coherent enhancement in the nucleus, increasing the rate for direct detection searches.   The spin-independent DM-nucleon cross section is given by
\begin{align}
\sigma_{\rm{SI}}^{N} = \frac{  \mu_N^2 }{\pi}  \frac{   g^{\prime \, 4}    B_{\chi}^2  }{ M_{Z^{\prime}}^4 } \,, \qquad \mu_N = \frac{   m_{\chi}   m_N }{   m_{\chi} + m_N } \,,
\end{align}
with $\mu_N$ being the DM-nucleon reduced mass.

The strongest limits on the spin-independent DM-nucleon cross section have been set by the Large Underground Xenon (LUX) experiment and by The Particle and Astrophysical Xenon Detector (PandaX).    The latest bounds set by the LUX experiment are based on a dataset taken from September 2014 to May 2016 (332 live days)~\cite{Akerib:2016vxi}.    The limits from the PandaX-II experiment are very similar and are based on a combination of the dataset taken during the commissioning run (19.1 live days) and the first physics run taking place from March 9 to June 30, 2016 (79.6 live days)~\cite{Tan:2016zwf}.   In Fig.~\ref{fig:plot2} we show exclusion limits on the model parameter space derived from these searches.   The limits are shown on the plane $m_{\chi} - M_{Z^{\prime}}$, fixing: $g^{\prime} = 0.4$, $\lambda^{\mu L} = 0.8$, $\tan \beta = 20$ and $B_{\chi}=1/3$.   We see that a large part of the allowed region left from flavour and collider bounds is excluded by direct DM searches.     For larger values of $B_{\chi}$, the derived constraints from direct searches basically exclude all the available parameter space shown in Fig.~\ref{fig:plot2}.

\section{Conclusions}  \label{sec:conc}

Motivated by the observed departures from the Standard Model in $b \to s \ell^+ \ell^-$ decays, we have performed a comprehensive analysis of a family-dependent $\mathrm{U(1)}^{\prime}$ model featuring a Dirac fermion dark matter candidate.  The simple model presented establishes a connection between dark matter stability and the hints of new physics in $b \to s \ell^+ \ell^-$, with the $Z^{\prime}$ boson acting as a portal between the dark sector and the Standard Model.

We found that the model can accommodate the current $b \to s \ell^+ \ell^-$ decay anomalies while satisfying flavour and collider constraints for a $Z^{\prime}$ boson in the TeV range.
For such a heavy mediator,
we rely on the Wigner enhancement effect to achieve the right size of the dark matter annihilation cross section
according to the current value of the dark matter relic abundance.
Thus in our model, the dark matter candidate $\chi$ is in TeV range  with $m_\chi \sim M_{Z'}/2$.
Fermi-LAT limits on dark matter annihilation from dwarf spheroidal galaxy observations exclude a very narrow, near-resonance, region of the model. In contrast, the latest limits from direct dark matter searches by the LUX and PandaX-II experiments
severely constrain the available model parameter space.

\appendix

\section{Details of the scalar sector}
\label{aop:sec}

The complex Higgs doublets $H_i$ can be parametrized as
\begin{equation}\label{eq:Phi}
H_i =  \begin{pmatrix}
H_i^+\\
H_i^0
\end{pmatrix}  =
\begin{pmatrix}
H^+_i\\
\frac{1}{\sqrt{2}}\left(v_i+\rho_i+i\eta_i\right)
\end{pmatrix}\quad(i=1,2)\,.
\end{equation}
The complex scalar field $\phi$ can be similarly parametrized in terms of its real components
\begin{equation}
\phi =  \frac{(    v_{\phi}+ R_0 + i I_0  )}{\sqrt{2}} \,.
\end{equation}
The scalar potential includes all terms invariant under the $\mathrm{SU(2)}_L \times \mathrm{U(1)}_Y \times \mathrm{U(1)}^{\prime}$ gauge symmetry.  The $\mathrm{U(1)}^{\prime}$ charge of  $\phi$ is constrained to be $X=\pm1$ or $X = \pm 1/2$ in order to avoid a Goldstone boson in the spectrum.    Additionally, we obtained that $X$ should be positive in order to accommodate $B \to K^{(*)} \ell^+ \ell^-$ data.    We choose here $X =  1$.      The scalar potential reads then
\begin{align}
V &=     m_i^2 |H_i|^2   + \frac{\lambda_i}{2}  |H_i|^4  + \lambda_3 |H_1|^2 |H_2|^2  \nonumber \\
&+ \lambda_4 | H_1^{\dag}  H_2 |^2  + \frac{b_2}{2}  |\phi|^2  + \frac{d_2}{4}  |\phi|^4 \nonumber \\
& + \frac{\delta_i}{2} |H_i|^2 |\phi|^2 -  \frac{\mu}{\sqrt{2}}  \left(  H_1^{\dag} H_2  \phi^*  + \mathrm{h.c.} \right)  \,.
\end{align}
We assume that $\mu \sim \mathcal{O}(v_{\phi})$.    After the gauge symmetry $\mathrm{SU(2)}_L \times \mathrm{U(1)}_Y \times \mathrm{U(1)}^{\prime}$ is spontaneously broken to the electromagnetic group, four degrees of freedom of the scalar fields give rise to the $W^{\pm}, Z^{(\prime)}$ masses.   The scalar spectrum contains three CP-even Higgs bosons $\{h_1,h_2,h_3\}$, a CP-odd Higgs $A$ and a charged scalar $H^{\pm}$.  We perform a perturbative diagonalization in the small ratio $v/v_{\phi}$ as in~\cite{Celis:2015ara}.    In terms of the neutral components of the scalar fields, the CP-even physical states are
\begin{equation}
\begin{pmatrix}
h_1\\
h_2\\
h_3
\end{pmatrix} \simeq \begin{pmatrix}
c_\beta&s_\beta   & 0 \\
-s_\beta&c_\beta  & 0  \\
0 & 0 & 1
\end{pmatrix}
\begin{pmatrix}
\rho_1\\
\rho_2 \\
R_0
\end{pmatrix} \,.
\end{equation}
Here we assumed a decoupling scenario, with $h_1$ a light SM-like Higgs boson to be identified with the $125$~GeV boson and additional heavy scalars around the scale $v_{\phi}$.   In the CP-odd sector, we have the would-be Goldstone bosons giving mass to the $Z$ and $Z^{\prime}$ bosons $G^{0 (\prime)}$ and the CP-odd Higgs $A$,
\begin{equation}
\begin{pmatrix}
G^0\\
A\\
G^{0 \prime}
\end{pmatrix} \simeq \begin{pmatrix}
c_\beta&s_\beta & 0   \\
-s_\beta&c_\beta   & 0  \\
0 & 0 & 1
\end{pmatrix}
\begin{pmatrix}
\eta_1\\
\eta_2 \\
I_0
\end{pmatrix} \,.
\end{equation}
The charged Higgs is given by
\begin{equation}
\begin{pmatrix}
G^{\pm}\\
H^{\pm}
\end{pmatrix} = \begin{pmatrix}
c_\beta&s_\beta   \\
-s_\beta&c_\beta
\end{pmatrix}
\begin{pmatrix}
H_1^{\pm}\\
H_2^{\pm}
\end{pmatrix} \,,
\end{equation}
with $G^{\pm}$ giving the $W^{\pm}$ mass.  The masses of the scalars are
\begin{align}
M_{h_2, A, H^{\pm}}^2 & \simeq  \frac{   \mu  v_{\phi}}{   s_{2 \beta} }   \,, \quad
 M_{h_3}^2 \simeq \frac{  d_2   v_{\phi}^2}{   2 }  \,, \quad
M_{h_1}^2 \simeq    \frac{ \tilde \lambda v^2 }{2 d_2}\,.
\end{align}
The combination $\tilde \lambda$ is given by
\begin{align}
\tilde \lambda =& \; c_{\beta}^4  (    2 d_2  \lambda_1 - \delta_1^2   )    + s_{\beta}^4 (     2 d_2 \lambda_2 - \delta_2^2 )  \nonumber \\
&+ 4 c_{\beta} s_{\beta}^3   \frac{\delta_2  \mu}{  v_{\phi} }   +  4 c_{\beta}^3  s_{\beta} \frac{  \delta_1 \mu  }{  v_{\phi} } \nonumber \\
&- 2 c_{\beta}^2 s_{\beta}^2 \left[     \delta_1 \delta_2   + 2 \left(    \frac{ \mu^2  }{  v_{\phi}^2}     - d_2 (  \lambda_3 + \lambda_4    )  \right)        \right]  \,.
\end{align}

\vspace{0.4cm}

\acknowledgments

The work of A.C. is supported by the Alexander von Humboldt Foundation.
A.C. is grateful to the Mainz Institute for Theoretical Physics (MITP), the Universit\`a di Napoli Federico II and INFN for its hospitality and its partial support during the completion of this work.
WZ.F. is grateful to Xiaoyong Chu, Zuowei Liu and Wei Xue for useful discussions.
The work of M.V. was supported by DFG Cluster of Excelence ``Origin and Structure of the Universe''. M.V. would like to thank the Theoretical Particle Physics group at the Ludwig-Maximilians-University of Munich and in particular Prof. Gerhard Buchalla's group for their hospitality while essential contributions to this work were completed.



\begin{thebibliography}{90}



\bibitem{Aaij:2014ora}
  R.~Aaij {\it et al.} [LHCb Collaboration],
  Phys.\ Rev.\ Lett.\  {\bf 113} (2014) 151601
 [arXiv:1406.6482 [hep-ex]].



\bibitem{Hiller:2003js}
  G.~Hiller and F.~Kruger,
  Phys.\ Rev.\ D {\bf 69} (2004) 074020
  [hep-ph/0310219].

\bibitem{Guevara:2015pza}
  A.~Guevara, G.~Lopez Castro, P.~Roig and S.~L.~Tostado,
  Phys.\ Rev.\ D {\bf 92} (2015) no.5,  054035
  [arXiv:1503.06890 [hep-ph]].

\bibitem{Bordone:2016gaq}
  M.~Bordone, G.~Isidori and A.~Pattori,
  arXiv:1605.07633 [hep-ph].



\bibitem{Alonso:2014csa}
  R.~Alonso, B.~Grinstein and J.~Martin Camalich,
  Phys.\ Rev.\ Lett.\  {\bf 113} (2014) 241802
  [arXiv:1407.7044 [hep-ph]].

\bibitem{Ghosh:2014awa}
  D.~Ghosh, M.~Nardecchia and S.~A.~Renner,
  JHEP {\bf 1412}, 131 (2014)
  [arXiv:1408.4097 [hep-ph]].

\bibitem{Jager:2014rwa}
  S.~J\"ager and J.~Martin Camalich,
  Phys.\ Rev.\ D {\bf 93} (2016) no.1,  014028
  [arXiv:1412.3183 [hep-ph]].

\bibitem{Altmannshofer:2015sma}
  W.~Altmannshofer and D.~M.~Straub,
  arXiv:1503.06199 [hep-ph].

\bibitem{Descotes-Genon:2015uva}
  S.~Descotes-Genon, L.~Hofer, J.~Matias and J.~Virto,
  JHEP {\bf 1606} (2016) 092
  [arXiv:1510.04239 [hep-ph]].

\bibitem{Ciuchini:2015qxb}
  M.~Ciuchini, M.~Fedele, E.~Franco, S.~Mishima, A.~Paul, L.~Silvestrini and M.~Valli,
  JHEP {\bf 1606} (2016) 116
  [arXiv:1512.07157 [hep-ph]].

\bibitem{Hurth:2016fbr}
  T.~Hurth, F.~Mahmoudi and S.~Neshatpour,
  Nucl.\ Phys.\ B {\bf 909} (2016) 737
  [arXiv:1603.00865 [hep-ph]].







\bibitem{Altmannshofer:2014cfa}
  W.~Altmannshofer, S.~Gori, M.~Pospelov and I.~Yavin,
  Phys.\ Rev.\ D {\bf 89} (2014) 095033
  [arXiv:1403.1269 [hep-ph]].

\bibitem{Crivellin:2015mga}
  A.~Crivellin, G.~D'Ambrosio and J.~Heeck,
  Phys.\ Rev.\ Lett.\  {\bf 114} (2015) 151801
  [arXiv:1501.00993 [hep-ph]].

\bibitem{Crivellin:2015lwa}
  A.~Crivellin, G.~D'Ambrosio and J.~Heeck,
  Phys.\ Rev.\ D {\bf 91} (2015) no.7,  075006
  [arXiv:1503.03477 [hep-ph]].

\bibitem{Sierra:2015fma}
  D.~Aristizabal Sierra, F.~Staub and A.~Vicente,
  Phys.\ Rev.\ D {\bf 92} (2015) no.1,  015001
  [arXiv:1503.06077 [hep-ph]].

\bibitem{Celis:2015ara}
  A.~Celis, J.~Fuentes-Martin, M.~Jung and H.~Serodio,
  Phys.\ Rev.\ D {\bf 92} (2015) no.1,  015007
  [arXiv:1505.03079 [hep-ph]].

\bibitem{Belanger:2015nma}
  G.~Belanger, C.~Delaunay and S.~Westhoff,
  Phys.\ Rev.\ D {\bf 92} (2015) 055021
  [arXiv:1507.06660 [hep-ph]].

\bibitem{Falkowski:2015zwa}
  A.~Falkowski, M.~Nardecchia and R.~Ziegler,
  JHEP {\bf 1511} (2015) 173
  [arXiv:1509.01249 [hep-ph]].

\bibitem{Allanach:2015gkd}
  B.~Allanach, F.~S.~Queiroz, A.~Strumia and S.~Sun,
  Phys.\ Rev.\ D {\bf 93} (2016) no.5,  055045
  [arXiv:1511.07447 [hep-ph]].

\bibitem{Celis:2015eqs}
  A.~Celis, W.~Z.~Feng and D.~L\"ust,
  JHEP {\bf 1602} (2016) 007
  [arXiv:1512.02218 [hep-ph]].





\bibitem{Chiang:2016qov}
  C.~W.~Chiang, X.~G.~He and G.~Valencia,
  Phys.\ Rev.\ D {\bf 93} (2016) no.7,  074003
  [arXiv:1601.07328 [hep-ph]].

\bibitem{Boucenna:2016wpr}
  S.~M.~Boucenna, A.~Celis, J.~Fuentes-Martin, A.~Vicente and J.~Virto,
  Phys.\ Lett.\ B {\bf 760} (2016) 214
  [arXiv:1604.03088 [hep-ph]].

\bibitem{Boucenna:2016qad}
  S.~M.~Boucenna, A.~Celis, J.~Fuentes-Martin, A.~Vicente and J.~Virto,
  arXiv:1608.01349 [hep-ph].


\bibitem{Jackson:2013pjq}
  C.~B.~Jackson, G.~Servant, G.~Shaughnessy, T.~M.~P.~Tait and M.~Taoso,
  JCAP {\bf 1307} (2013) 021
  [arXiv:1302.1802 [hep-ph]].

\bibitem{Jackson:2013rqp}
  C.~B.~Jackson, G.~Servant, G.~Shaughnessy, T.~M.~P.~Tait and M.~Taoso,
  JCAP {\bf 1307} (2013) 006
  [arXiv:1303.4717 [hep-ph]].

\bibitem{Duerr:2013lka}
  M.~Duerr and P.~Fileviez Perez,
  Phys.\ Lett.\ B {\bf 732} (2014) 101
  [arXiv:1309.3970 [hep-ph]].

\bibitem{Alves:2013tqa}
  A.~Alves, S.~Profumo and F.~S.~Queiroz,
  JHEP {\bf 1404} (2014) 063
  [arXiv:1312.5281 [hep-ph]].

\bibitem{Arcadi:2013qia}
  G.~Arcadi, Y.~Mambrini, M.~H.~G.~Tytgat and B.~Zaldivar,
  JHEP {\bf 1403} (2014) 134
  [arXiv:1401.0221 [hep-ph]].

\bibitem{Feng:2014cla}
  W.~Z.~Feng, G.~Shiu, P.~Soler and F.~Ye,
  JHEP {\bf 1405}, 065 (2014)
  [arXiv:1401.5890 [hep-ph]].

\bibitem{Bell:2014tta}
  N.~F.~Bell, Y.~Cai, R.~K.~Leane and A.~D.~Medina,
  Phys.\ Rev.\ D {\bf 90} (2014) no.3,  035027
  [arXiv:1407.3001 [hep-ph]].

\bibitem{Duerr:2014wra}
  M.~Duerr and P.~Fileviez Perez,
  Phys.\ Rev.\ D {\bf 91} (2015) no.9,  095001
  [arXiv:1409.8165 [hep-ph]].

\bibitem{Hooper:2014fda}
  D.~Hooper,
  Phys.\ Rev.\ D {\bf 91} (2015) 035025
  [arXiv:1411.4079 [hep-ph]].

\bibitem{Alves:2015pea}
  A.~Alves, A.~Berlin, S.~Profumo and F.~S.~Queiroz,
  Phys.\ Rev.\ D {\bf 92} (2015) no.8,  083004
  [arXiv:1501.03490 [hep-ph]].

\bibitem{Duerr:2015wfa}
  M.~Duerr, P.~Fileviez Perez and J.~Smirnov,
  Phys.\ Rev.\ D {\bf 92} (2015) no.8,  083521
  [arXiv:1506.05107 [hep-ph]].

\bibitem{Alves:2015mua}
  A.~Alves, A.~Berlin, S.~Profumo and F.~S.~Queiroz,
  JHEP {\bf 1510} (2015) 076
  [arXiv:1506.06767 [hep-ph]].



\bibitem{Babu:1997st}
  K.~S.~Babu, C.~F.~Kolda and J.~March-Russell,
  Phys.\ Rev.\ D {\bf 57} (1998) 6788
  [hep-ph/9710441].


\bibitem{Meinel:2016grj}
  S.~Meinel and D.~van Dyk,
  Phys.\ Rev.\ D {\bf 94}, no. 1, 013007 (2016)
  [arXiv:1603.02974 [hep-ph]].


\bibitem{Aad:2014cka}
  G.~Aad {\it et al.} [ATLAS Collaboration],
  Phys.\ Rev.\ D {\bf 90}, no. 5, 052005 (2014)
  [arXiv:1405.4123 [hep-ex]].


\bibitem{Khachatryan:2014fba}
  V.~Khachatryan {\it et al.} [CMS Collaboration],
  JHEP {\bf 1504} (2015) 025
  [arXiv:1412.6302 [hep-ex]].



\bibitem{ATLAS:2016ichep}
  The ATLAS collaboration,
 ATLAS-CONF-2016-045.





\bibitem{Alwall:2014hca}
  J.~Alwall {\it et al.},
  JHEP {\bf 1407} (2014) 079
 [arXiv:1405.0301 [hep-ph]].

\bibitem{CMS:2012rya}
  CMS Collaboration,
  CMS-PAS-EXO-11-092.

\bibitem{Falkowski:2013jya}
  A.~Falkowski, D.~M.~Straub and A.~Vicente,
  JHEP {\bf 1405} (2014) 092
  [arXiv:1312.5329 [hep-ph]].

\bibitem{Dermisek:2014qca}
  R.~Dermisek, J.~P.~Hall, E.~Lunghi and S.~Shin,
  JHEP {\bf 1412}, 013 (2014)
  [arXiv:1408.3123 [hep-ph]].

\bibitem{Lenz:2010gu}
  A.~Lenz {\it et al.},
  Phys.\ Rev.\ D {\bf 83} (2011) 036004
  [arXiv:1008.1593 [hep-ph]].

\bibitem{Buras:2012dp}
  A.~J.~Buras, F.~De Fazio, J.~Girrbach and M.~V.~Carlucci,
  JHEP {\bf 1302} (2013) 023
  [arXiv:1211.1237 [hep-ph]].

\bibitem{Buras:2012jb}
  A.~J.~Buras, F.~De Fazio and J.~Girrbach,
  JHEP {\bf 1302} (2013) 116
  [arXiv:1211.1896 [hep-ph]].

\bibitem{CKMgroup}
CKMfitter Group (J. Charles et al.), Eur. Phys. J. C41, 1-131 (2005) [hep-ph/0406184], updated results and plots available at: http://ckmfitter.in2p3.fr

\bibitem{1602.03560}
  A.~Bazavov {\it et al.} [Fermilab Lattice and MILC Collaborations],
  arXiv:1602.03560 [hep-lat].






\bibitem{1412.7515}
  Y.~Amhis {\it et al.} [Heavy Flavor Averaging Group (HFAG) Collaboration],
  arXiv:1412.7515 [hep-ex].


\bibitem{Hiller:2014ula}
  G.~Hiller and M.~Schmaltz,
  JHEP {\bf 1502} (2015) 055
  [arXiv:1411.4773 [hep-ph]].


\bibitem{Buras:2014fpa}
  A.~J.~Buras, J.~Girrbach-Noe, C.~Niehoff and D.~M.~Straub,
  JHEP {\bf 1502} (2015) 184
  doi:10.1007/JHEP02(2015)184
  [arXiv:1409.4557 [hep-ph]].




\bibitem{Lees:2013kla}
  J.~P.~Lees {\it et al.} [BaBar Collaboration],
  Phys.\ Rev.\ D {\bf 87} (2013) no.11,  112005
  doi:10.1103/PhysRevD.87.112005
  [arXiv:1303.7465 [hep-ex]].

\bibitem{Lutz:2013ftz}
  O.~Lutz {\it et al.} [Belle Collaboration],
  Phys.\ Rev.\ D {\bf 87} (2013) no.11,  111103
  doi:10.1103/PhysRevD.87.111103
  [arXiv:1303.3719 [hep-ex]].

\bibitem{Hiller:2014yaa}
  G.~Hiller and M.~Schmaltz,
  Phys.\ Rev.\ D {\bf 90} (2014) 054014
  [arXiv:1408.1627 [hep-ph]].


\bibitem{CMS:2014xfa}
  V.~Khachatryan {\it et al.} [CMS and LHCb Collaborations],
  Nature {\bf 522} (2015) 68
  [arXiv:1411.4413 [hep-ex]].



\bibitem{Bobeth:2013uxa}
  C.~Bobeth, M.~Gorbahn, T.~Hermann, M.~Misiak, E.~Stamou and M.~Steinhauser,
  Phys.\ Rev.\ Lett.\  {\bf 112} (2014) 101801
  [arXiv:1311.0903 [hep-ph]].


\bibitem{Aaboud:2016ire}
  M.~Aaboud {\it et al.} [ATLAS Collaboration],
  arXiv:1604.04263 [hep-ex].

\bibitem{Weinberg:1979sa}
  S.~Weinberg,
  Phys.\ Rev.\ Lett.\  {\bf 43} (1979) 1566.



\bibitem{Griest:1990kh}
  K.~Griest and D.~Seckel,
  Phys.\ Rev.\ D {\bf 43}, 3191 (1991).

\bibitem{Gondolo:1990dk}
  P.~Gondolo and G.~Gelmini,
  Nucl.\ Phys.\ B {\bf 360}, 145 (1991).


\bibitem{Nath:1992ty}
  P.~Nath and R.~L.~Arnowitt,
  Phys.\ Rev.\ Lett.\  {\bf 70}, 3696 (1993)
  [hep-ph/9302318].

\bibitem{Ibe:2008ye}
  M.~Ibe, H.~Murayama and T.~T.~Yanagida,
  Phys.\ Rev.\ D {\bf 79}, 095009 (2009)
  [arXiv:0812.0072 [hep-ph]].



\bibitem{Ackermann:2015zua}
  M.~Ackermann {\it et al.} [Fermi-LAT Collaboration],
  Phys.\ Rev.\ Lett.\  {\bf 115} (2015) no.23,  231301
  [arXiv:1503.02641 [astro-ph.HE]].


\bibitem{Ahnen:2016qkx}
  M.~L.~Ahnen {\it et al.} [MAGIC and Fermi-LAT Collaborations],
  JCAP {\bf 1602} (2016) no.02,  039
  [arXiv:1601.06590 [astro-ph.HE]].


\bibitem{::2016jja}
  H.~Abdallah {\it et al.} [HESS Collaboration],
  arXiv:1607.08142 [astro-ph.HE].


\bibitem{Undagoitia:2015gya}
  T.~Marrodán Undagoitia and L.~Rauch,
  J.\ Phys.\ G {\bf 43} (2016) no.1,  013001
  [arXiv:1509.08767 [physics.ins-det]].



\bibitem{D'Eramo:2016atc}
  F.~D'Eramo, B.~J.~Kavanagh and P.~Panci,
  arXiv:1605.04917 [hep-ph].

 \bibitem{runDM}
  F.~D'Eramo, B. J. Kavanagh and P. Panci (2016). runDM (Version 1.0) [Computer software].  Available at \url{https://github.com/bradkav/runDM/}

\bibitem{Akerib:2016vxi}
  D.~S.~Akerib {\it et al.} [LUX Collaboration],
  Phys.\ Rev.\ Lett.\  {\bf 118} (2017) no.2,  021303
  doi:10.1103/PhysRevLett.118.021303
  [arXiv:1608.07648 [astro-ph.CO]].

\bibitem{Tan:2016zwf}
  A.~Tan {\it et al.} [PandaX-II Collaboration],
  arXiv:1607.07400 [hep-ex].

\end{thebibliography}
\end{document}